\documentclass[journal=jpcafh,manuscript=article]{achemso} 
\usepackage[version=3]{mhchem} 
\usepackage{caption}
\usepackage{subcaption}
\usepackage{color}
\usepackage{xcolor}
\usepackage{comment}
\usepackage{amssymb}

\usepackage{soul}

\author{Hongseok Choi}
\altaffiliation{Contributed equally to this work}
\affiliation{Department of Chemistry, Korea Advanced Institute of Science and Technology (KAIST), Daejeon 34141, Korea}
\author{Kyungmin Kim}
\altaffiliation{Contributed equally to this work}
\affiliation{Department of Chemistry, Korea Advanced Institute of Science and Technology (KAIST), Daejeon 34141, Korea}
\author{Young Min Rhee}
\affiliation{Department of Chemistry, Korea Advanced Institute of Science and Technology (KAIST), Daejeon 34141, Korea}
\email{ymrhee@kaist.ac.kr}

\title
  {Iterative Projection-Based Embedding Scheme Combined with Variational Quantum Eigensolver}

\begin{document}

%
%

\begin{abstract}

Quantum embedding methods offer a promising route to extend quantum chemical calculations to large multiscale systems by treating a chemically important subsystem at a high level of theory while describing its surrounding environment at an affordable level. The methods are also quite relevant for quantum computing approaches based on hardware with limited resources. Here, we present an iterative projection-based embedding framework combined with the variational quantum eigensolver (VQE), in which the environment density is allowed to respond self-consistently to the refined electronic structure of the embedded subsystem described by VQE. Unlike conventional ``one-shot'' embedding approaches where the environment remains frozen after the initial orbital optimization, the proposed iterative scheme alternates between the VQE-level treatment of the subsystem and a mean-field-level refinement of the environment until mutual self-consistency is achieved. The convergence behavior of the iterative scheme is first examined using several small test systems. Its practical applicability is then demonstrated with a composite system consisting of a methylenimine (CH$_2$NH) molecule sandwiched between two benzene rings, with the H$-$C=N$-$H dihedral angle rotating from 0$^{\circ}$ to 90$^{\circ}$. The iterative procedure consistently converges within $\sim$10 macro-iteration steps across all tested geometries, yielding energies below the conventional one-shot embedding results. The converged results well reproduce the fully correlated reference energy employing the same active space, and the resulting potential energy surface with respect to the dihedral rotation is also in good agreement with the reference one. These results demonstrate that our iterative embedding framework is numerically robust and physically sound, yielding a self-consistent and reliable treatment of inter-subsystem correlation. We expect that its formulation will be particularly compatible with the emerging paradigm of quantum-classical hybrid computing.


\end{abstract}

\newpage

\section{1. Introduction}

Computation of physical and chemical processes often requires access to the electronic structures of the systems involved, but treating complex multiscale ones often suffers from the expensive nature of the theory. To balance between accuracy and efficiency, accordingly, numerous embedding style computational tactics have been developed. Conventionally, an embedding method focuses on a region of significance by adopting a high-level calculation while handling the remaining environment through an inexpensive and more affordable theory. In a broad sense, the quantum mechanics / molecular mechanics (QM/MM) approaches,\cite{Warshel1976, Field1990, Senn2009} their related ONIOM methods,\cite{Vreven2006, Vreven2006b, Chung2015} and frozen-density embedding (FDE)\cite{cortona1991self,wesolowski1993frozen} together with its later developments into subsystem density functional theory \cite{Trail2000, Jacob2014, wesolowski2015frozen, hofener2012molecular} can be considered representative examples. Related divide-and-conquer strategies, such as fragment-based methods\cite{Fedorov2007,gordon2012fragmentation} and density matrix embedding theory,\cite{Knizia2013,wouters2016practical,giordano2026ab} share a similar philosophy of partitioning a large system into smaller, more tractable subsystems.

Among these, the projection-based embedding scheme\cite{Manby2012, Lee2019a, Huo2016, Bennie2016, Zhang2018, Barnes2015, Bennie2017, Welborn2018, bensberg2020orbital, chulhai2018projection} provides a transparent framework in which a subsystem of interest is described by a high-level wave function theory (WFT), while its interaction with the surrounding environment is retained at the self-consistent-field (SCF) level. This approach allows one to systematically incorporate accurate electronic structure methods for the selected region without incurring the prohibitive cost of a full high-level treatment of the entire system. Owing to this favorable balance between accuracy and efficiency, projection-based embedding has been successfully applied to a variety of large-scale problems, including transition metal catalysis,\cite{Huo2016} bio-enzyme catalysis,\cite{Bennie2016, Zhang2018} electron transfer processes in complex environments,\cite{Barnes2015, Bennie2016} excitation energies,\cite{Bennie2017} reaction pathway calculation,\cite{Welborn2018, bensberg2020orbital} and periodic solid-state systems.\cite{chulhai2018projection}

The coupled cluster (CC) theory has been widely adopted within the projection-based embedding framework as the high-level method.\cite{Bennie2016, Bennie2017, Zhang2018} However, the nonvariational nature of conventional CC methods\cite{bartlett2007coupled} can occasionally lead to unphysical results, motivating the search for more robust alternatives. The unitary coupled cluster (UCC) approach\cite{bartlett1989alternative, taube2006new} may address this issue as its unitarity guarantees a variational upper bound for the ground-state energy. 
In addition, UCC has become particularly compelling with the advent of quantum computing hardware,\cite{Peruzzo2014, anand2022quantum} as the unitary structure of the ansatz maps naturally onto quantum circuit operations. The UCC ansatz has therefore been naturally combined with variational quantum algorithms such as the variational quantum eigensolver (VQE),\cite{cerezo2021variational, tilly2022variational} which has become a practical framework for quantum-classical hybrid electronic structure calculations.\cite{Peruzzo2014, mcclean2016theory, o2016scalable} Together with its relatively modest circuit-depth requirements, VQE is particularly well suited to noisy intermediate-scale quantum (NISQ) devices. Consequently, numerous VQE ansätze and parameterized quantum circuit constructions have been proposed, including both UCC-based approaches and alternative hardware-efficient or problem-inspired formulations.\cite{Huggins2020a, Fan2023a, Grimsley2020a, Tang2021, Nakagawa2023a, Yalouz2022a, Mizukami2020, Romero2018a, Omiya2022, Bierman2023a, barkoutsos2018quantum, Kandala2017, Xia2020a, gard2020efficient, ryabinkin2018qubit} Consequently, VQE calculations for small molecular systems have been successfully demonstrated on both quantum simulators and quantum hardware.\cite{o2016scalable, google2020hartree}
More recently, VQE-based methodologies have been extended toward the treatment of larger multiscale systems, and the embedding approach in the manner of VQE-in-SCF has been successfully demonstrated.\cite{Rossmannek2023, Ralli2024}


Despite this favorable balance between accuracy and efficiency, the conventional projection-based embedding framework typically adopts a frozen description of the environment, in which the environment electron density is fixed at the lower-level SCF solution.\cite{Manby2012, Lee2019a} While this approximation can be adequate when the subsystem--environment coupling is weak, it may introduce systematic errors when the electronic structure of the environment is significantly coupled to that of the embedded subsystem. In such cases, simply replacing the embedded subsystem treatment with a higher-level VQE description is not sufficient on its own. Namely, the environment density should also respond self-consistently to the updated electronic structure of the embedded region. This aspect naturally motivates an iterative embedding formulation, in which the environment is allowed to relax in response to the VQE-level treatment of the subsystem. This consideration becomes particularly important in VQE-in-SCF approaches, where VQE is specifically introduced to treat embedded regions whose electronic structures deviate substantially from a mean-field description.

Similar self-consistent embedding concepts have previously been explored within the FDE framework, where the freeze-and-thaw procedure iteratively relaxes the environment density until self-consistency.\cite{wesolowski1993frozen, wesolowski1996kohn, jacob2008flexible, wesolowski2015frozen} However, FDE inherently relies on approximate kinetic-energy density functionals to evaluate the nonadditive kinetic potential, introducing an additional source of error when the subsystem densities overlap significantly. In contrast, projection-based embedding eliminates the need for such approximations by enforcing orbital orthogonality through a level-shift projection operator, rendering the nonadditive kinetic potential exactly zero.\cite{Manby2012} Despite this advantage, projection-based embedding has thus far generally retained a frozen description of the environment during the high-level calculation, and a self-consistent environment relaxation procedure analogous to freeze-and-thaw has not yet been established within the VQE-in-SCF framework.

In this work, we try to address this limitation by introducing iterative environment relaxation into the projection-based VQE-in-SCF embedding framework.
We extend the ``\text{A}-in-\text{B}'' framework by introducing an iterative environment response, enabling a self-consistent treatment of subsystem--environment coupling beyond the conventional one-shot approximation. Specifically, rather than keeping the environment density frozen at the initial SCF solution throughout the VQE calculation, we propose a scheme with an added layer of iterations. Because the one-particle density can be readily measured from the VQE side, the environment density is updated via an embedded ``\text{B}-in-\text{A}'' SCF procedure at each iteration, allowing the environment to respond to the refined electronic structure of the embedded subsystem. This mutual adaptation is repeated until both the total energy and the environment density converge to self-consistency, yielding a fully consistent description of the embedded electronic structure. To ensure numerical stability of the iterative procedure, a damping scheme is incorporated into the density update, providing robust convergence across a range of subsystem--environment coupling strengths.

\section{2. Methods}
\subsection{2.1. VQE-in-SCF embedding}
While the details of the projection-based embedding method can be found elsewhere,\cite{Lee2019a, Barnes2013, Goodpaster2014, Lee2019b, Manby2012} we briefly review its formulation here for the sake of completeness. The method starts with an SCF calculation, such as Hartree--Fock (HF) or density functional theory (DFT), for the whole system, and the molecular orbitals (MOs) thus obtained are spatially localized and partitioned into an embedded subsystem and an embedding environment based on predefined groups of atoms. For simplicity and consistency with the existing literature,\cite{Manby2012} we denote the embedded subsystem and the embedding environment as $\text{A}$ and $\text{B}$, respectively. Subsystem $\text{A}$ represents the region of primary interest and is subsequently treated at a higher level of theory.
Various orbital localization schemes can be employed to localize the orbitals to A and B, including the Boys\cite{boys1960construction, foster1960canonical} and the Pipek-Mezey methods\cite{pipek1989fast}, where the logic is based on the absolute distances or the Mulliken populations, as well as the subsystem projected atomic orbital decomposition (SPADE) procedure,\cite{Claudino2019a} where singular value decomposition directly separates the orbital space into the two subsystems.

After partitioning, the two sets of orbitals are used to construct the density matrices $\gamma^\text{A}$ and $\gamma^\text{B}$. The low-level density matrix functional\cite{Manby2012} of two-electron terms is defined as
\begin{equation}
    \mathbf{g}[\gamma]
    =
    \mathbf{J}[\gamma]
    +
    \mathbf{v}^{\text{xc}}[\gamma],
    \label{eq:g}
\end{equation}
where $\mathbf{J}[\gamma]$ denotes the Coulomb matrix and $\mathbf{v}^{\text{xc}}[\gamma]$ the exchange--correlation potential matrix evaluated for the density $\gamma$. Of course, the above formulation can be trivially extended for handling HF and hybrid DFT. Let us additionally denote a high-level density matrix functional as $\mathbf g'[\gamma]$. In SCF-in-SCF calculations, the low- and the high-level functionals differ only in the exchange--correlation functional, whereas the Coulomb contribution is identical.

To enforce mutual orthogonality between the occupied orbitals of subsystems
$\text{A}$ and $\text{B}$, projection-based embedding introduces the projection operator
\begin{equation}
\mathbf{P}^\text{B} = \mathbf{S} \gamma^\text{B}\mathbf{S},
\end{equation}
where
$\mathbf S$ is the atomic orbital (AO) overlap matrix with elements
$S_{\lambda\sigma}=\langle\chi_\lambda|\chi_\sigma\rangle$, together with a level-shift parameter $\mu$.
For sufficiently large $\mu$, the projection operator excludes the occupied
orbital space of subsystem B from the variational optimization of subsystem A,
thereby eliminating the need to explicitly treat the nonadditive kinetic-energy
contribution.\cite{Manby2012}

With these definitions, the embedded one-electron Hamiltonian is given by
\begin{eqnarray}
    \mathbf{h}^\text{A-in-B} &=& \mathbf{h} + \mathbf{g}[\gamma^\text{A} + \gamma^\text{B}] - \mathbf{g}[\gamma^\text{A}]
 + \mu \mathbf{P}^\text{B} 
 \nonumber \\
 &=& \mathbf{h} +\mathbf{J[\gamma^\text{B}]} + \mathbf{v}^\text{xc}[\gamma^\text{A} + \gamma^\text{B}] - \mathbf{v}^\text{xc}[\gamma^\text{A}] + \mu \mathbf{P}^\text{B} ,
 \label{eq:h_emb1}
\end{eqnarray}
where $\mathbf h$ denotes the conventional one-electron Hamiltonian containing the kinetic energy and electron--nuclear attraction terms.
Using the embedded one-electron Hamiltonian, the embedded Fock matrix is given by
\begin{equation}
    \mathbf{f}^\text{A}_\text{emb} \equiv \mathbf{h}^\text{A-in-B} + \mathbf{g}^\prime [\gamma^\text{A}_\text{emb}] ,
    \label{eq:fAemb}
\end{equation}
where $\gamma^\text A_\text{emb}$ is obtained self-consistently from the occupied eigenvectors of $\mathbf f^\text A_\text{emb}$.



The total energy at the initial SCF level can be partitioned into the energies of the two subsystems together with their nonadditive interaction,
\begin{equation}
    E= E^\text A[\gamma^\text A] + E^\text B[\gamma^\text B] + E_\text{nad}^{\text{A-B}}[\gamma^\text A,\gamma^\text B].
\end{equation}
When the subsystem $\text{A}$ is treated at a different level of theory, its density relaxes from the reference density $\gamma^\text A$ to $\gamma^\text A_\text{emb}$. Consequently, both the subsystem energy and the nonadditive interaction must be updated.
%
The total energy is therefore formally written as
\begin{equation}
    E_\text{SCF-in-SCF} = E^\text{A}[\gamma_\text{emb}^\text{A}] + E^\text{B}[\gamma^\text{B}] + E^\text{A-B}_\text{nad}[\gamma_\text{emb}^\text{A}, \gamma^\text{B}] .
    \label{eq:E_SCF-in-SCF}
\end{equation}
The dependence of $E^\text{A-B}_\text{nad}$ on the relaxed density $\gamma^\text{A}_\text{emb}$ is generally unknown. Following the original projection-based embedding formulation, the nonadditive interaction is approximated by a first-order expansion around the reference density $\gamma^\text{A}$.
This gives
\begin{equation}
    E_\text{SCF-in-SCF} = E^\text{A}[\gamma_\text{emb}^\text{A}] + \text{Tr} (\gamma_\text{emb}^\text{A} - \gamma^\text{A}) (\mathbf{h}^\text{A-in-B} - \mathbf{h})
    + E^\text{B}[\gamma^\text{B}] + E_\text{nad}^\text{A-B}[\gamma^\text{A}, \gamma^\text{B}] ,
    \label{eq:E_SCF-in-SCF2}
\end{equation}
where the second term represents the first-order correction arising from the density relaxation of subsystem A in the embedding potential.

When subsystem A is described by a wavefunction method, its energy is evaluated directly from the embedded Hamiltonian. Accordingly, the subsystem energy together with the corresponding first-order interaction correction can be replaced by the expectation value of the embedded Hamiltonian:
\begin{equation}
    E = \langle \Psi^\text{A} |\hat H^\text{A-in-B}| \Psi^\text{A}\rangle
     - \text{Tr} \gamma^\text{A} (\mathbf{h}^\text{A-in-B} - \mathbf{h})
    + E^\text{B}[\gamma^\text{B}] + E_\text{nad}^\text{A-B}[\gamma^\text{A}, \gamma^\text{B}] ,
    \label{eq:E_WFT-in-SCF}
\end{equation}
where the second term removes the reference interaction energy already contained in the embedded Hamiltonian, thereby avoiding double counting.

The VQE-in-SCF formulation is obtained by replacing the classical wavefunction solver in Eq.~(\ref{eq:E_WFT-in-SCF}) with a variational quantum eigensolver, while leaving the embedding framework unchanged:
\begin{equation}
\hat H^\text{A-in-B} = \sum_{pq} h_{pq}^{\text{A-in-B}} a_p^\dagger a_q
+ \frac12 \sum_{pqrs} \langle pq|rs\rangle a_p^\dagger a_q^\dagger a_s a_r.
\end{equation}
After the VQE optimization, the one-particle density matrix $\gamma^\text A_\text{emb}$ can be obtained from measurements on the optimized quantum state. This density is subsequently used in the iterative embedding procedure described in the following section.

\subsection{2.2. Iterative embedding}

Conventional projection-based embedding keeps the environment density fixed at the value obtained from the initial whole-system SCF calculation. As already explained, while this approximation is often effective, it cannot account for the response of the environment to changes in the embedded subsystem. This limitation will become particularly important in a regime where VQE becomes useful for treating subsystem A for its substantial deviation from the mean-field result. To address this limitation, we propose an iterative embedding scheme in which the embedded subsystem and its environment are updated alternately until self-consistency is achieved. In addition to the conventional embedding iterations (hereafter referred to as micro-iterations), we introduce an outer self-consistency cycle, referred to as the macro-iteration, to allow the environment density to relax in response to the updated subsystem density. Its scheme works as follows:

\textbf{Step 1: A-in-B as VQE-in-SCF calculation.} 
Using the current embedded Hamiltonian
$\hat H^\text{A-in-B}_i$, a VQE calculation is performed to obtain the embedded density matrix $\gamma^\text{A}_{\text{emb},i} \equiv \gamma^\text{A}_{\text{VQE},i}$. The resulting density matrix is transformed to the AO basis for use in the subsequent environment update.
 
\textbf{Step 2: B-update as refinement on environment.} 
Keeping the density in the above step fixed, we can then perform an SCF-in-SCF calculation with the roles of subsystems \text{A} and \text{B} interchanged to re-optimize subsystem B. Practically, we found that the macro-iterations can be more stable if the embedded density is mixed with the density from the previous macro-iteration step before updating the environment. Namely, \begin{equation}
\gamma^\text{A}_{i} = (1-\alpha)\gamma^\text{A}_{\text{emb},i} + \alpha\gamma^\text{A}_{\text{emb},i-1}
\end{equation}
with a damping factor $\alpha\in[0,1)$ is adopted as the embedded density for updating the description of B.

Using the damped density $\gamma^\text{A}_{i}$, the embedded core Hamiltonian for subsystem \text{B} is constructed as
\begin{equation}
    \mathbf h^\text{B-in-A}_{i} = \mathbf h + \mathbf g[\gamma^\text A_{i}+\gamma^\text B_i] - \mathbf g[\gamma^\text B_i] + \mu\mathbf P^\text A_{i},
\end{equation}
where the projector is $\mathbf P^\text A_{i} = \mathbf S \gamma^\text A_{i} \mathbf S$,
and the corresponding embedded Fock matrix is iteratively constructed as
$
    \mathbf f^\text B_{i} = 
    \mathbf h^\text{B-in-A}_{i} + \mathbf g[\gamma^\text B_{i}]
$.
The embedded SCF equations are solved self-consistently to obtain the updated environment density $\gamma^\text B_{i}$.

\textbf{Step 3: Reconstructing A-in-B Hamiltonian.} With the updated environment density $\gamma^\text B_{i}$, the embedded one-electron Hamiltonian for subsystem \text{A} is reconstructed as
\begin{equation}
   \mathbf{h}^\text{A-in-B}_{i} = \mathbf{h} + \mathbf{g}[\gamma^\text{A}_{i}+\gamma^\text{B}_{i}] - \mathbf{g}[\gamma^\text{A}_{i}] + \mu\mathbf{P}^\text{B}_{i} .
\end{equation}
The reconstructed Hamiltonian serves as the embedded Hamiltonian for the subsequent macro-iteration.
These three steps are repeated until the macro-iterations converge by monitoring both the total energy and the environment density. Upon convergence, the subsystem densities and the embedded Hamiltonian become mutually consistent.

\subsection{2.3. Demonstration details}
The projection-based embedding was implemented with the aid of PySCF,\cite{Sun2015, Sun2018, Sun2020} a quantum chemical computational package in Python. For the molecular orbital partitioning, three localization schemes were tested: the SPADE,\cite{Claudino2019a} the Boys,\cite{boys1960construction, foster1960canonical} and the Pipek–Mezey\cite{pipek1989fast} methods, all implemented within PySCF. All other classical reference computations, except for complete active space SCF (CASSCF) calculations, were also performed using PySCF. The CASSCF calculations were performed using Molpro.\cite{werner2020molpro}
The quantum circuit simulations, VQE in this case, were performed with the help of Qiskit.\cite{qiskit2024}. The quantum simulations in this work did not consider any statistical noise, and the expectation values were sampled using the statevector simulator. The unitary coupled-cluster with singles and doubles (UCCSD) ansatz\cite{romero2019strategies} was employed for VQE
using the STO-3G basis set. For all tested molecular system, the active space was selected from the localized occupied MOs and their corresponding virtual MOs so that each of the resulting problems required at most 16 qubits for the sake of resource economy. More specific details will be provided in the next section.

\section{3. Results and Discussion}
\subsection{3.1 Numerical Stability of the Iterative Embedding}
\begin{figure}[!ht]
    \centering
    \includegraphics[width=0.4\textwidth]{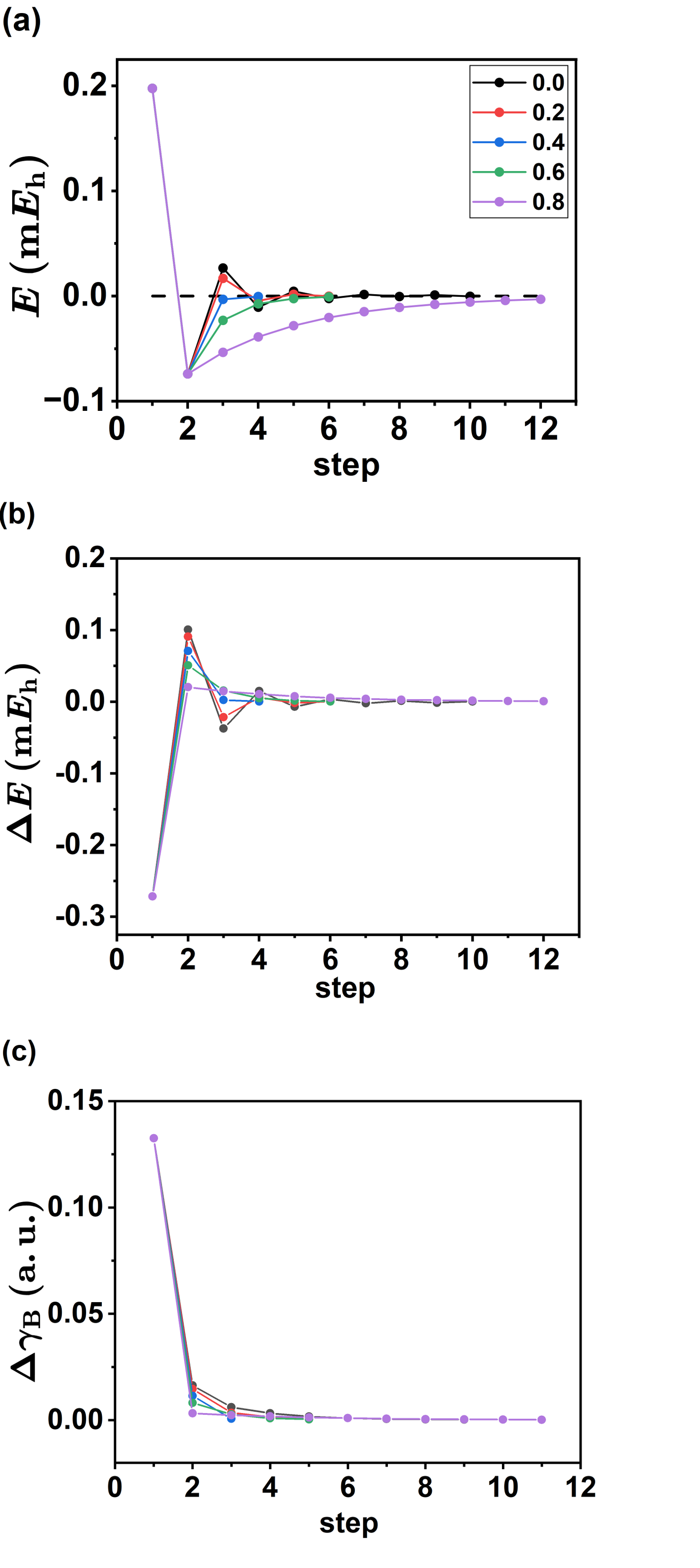}

    \caption{Convergence of the iterative embedding for a water dimer system with varying damping factors $\alpha$: (a) the energy difference from the converged value, as a function of the macro-iteration step, with the converged value set to zero (dashed line), (b) the changes of the total energy at each successive macro-iteration step, $\Delta E = E_{i+1} - E_i$, and (c) the stepwise environmental density change $\Delta\gamma_\text{B} = \|\gamma_\text{B}^{(i+1)} - \gamma_\text{B}^{(i)}\|_F / \mathrm{dim}(\gamma_\text{B})$, with $\| \cdot \|_F$ denoting Frobenius norm defined as $ \|A\|^2_F=\sum_{i,j} |A_{ij}|^2$.}
    \label{fig:waterdimer_iteration_embedding}
\end{figure}

We first examine the convergence behavior of the proposed iterative embedding method on a water dimer system, where subsystem A consists of one water monomer treated at the VQE level and subsystem B consists of the other water monomer treated at the HF level. The calculations were performed with varying damping factors ($\alpha = 0.0$, 0.2, 0.4, 0.6, and 0.8) to assess the robustness of the iterative procedure in relation to damping. Figure \ref{fig:waterdimer_iteration_embedding}a displays the energy results, and we can see that all calculations have converged to the same total energy regardless of $\alpha$, although the convergence pathways are somewhat different. Interestingly, but quite understandably, larger damping factors lead to slower but more monotonic convergence, as seen most clearly for $\alpha = 0.8$. The convergence is further confirmed by the stepwise energy difference $\Delta E$ shown in Figure \ref{fig:waterdimer_iteration_embedding}b as well as the environmental density change $\Delta\gamma_\text{B}$ shown in Figure \ref{fig:waterdimer_iteration_embedding}c, both of which reach convergence within $\sim$10 macro-iteration steps for all values of $\alpha$. These results demonstrate that the iterative procedure converges to a practically unique solution regardless of the damping parameter.

\begin{figure}[!ht]
    \centering
    \includegraphics[width=0.4\textwidth]{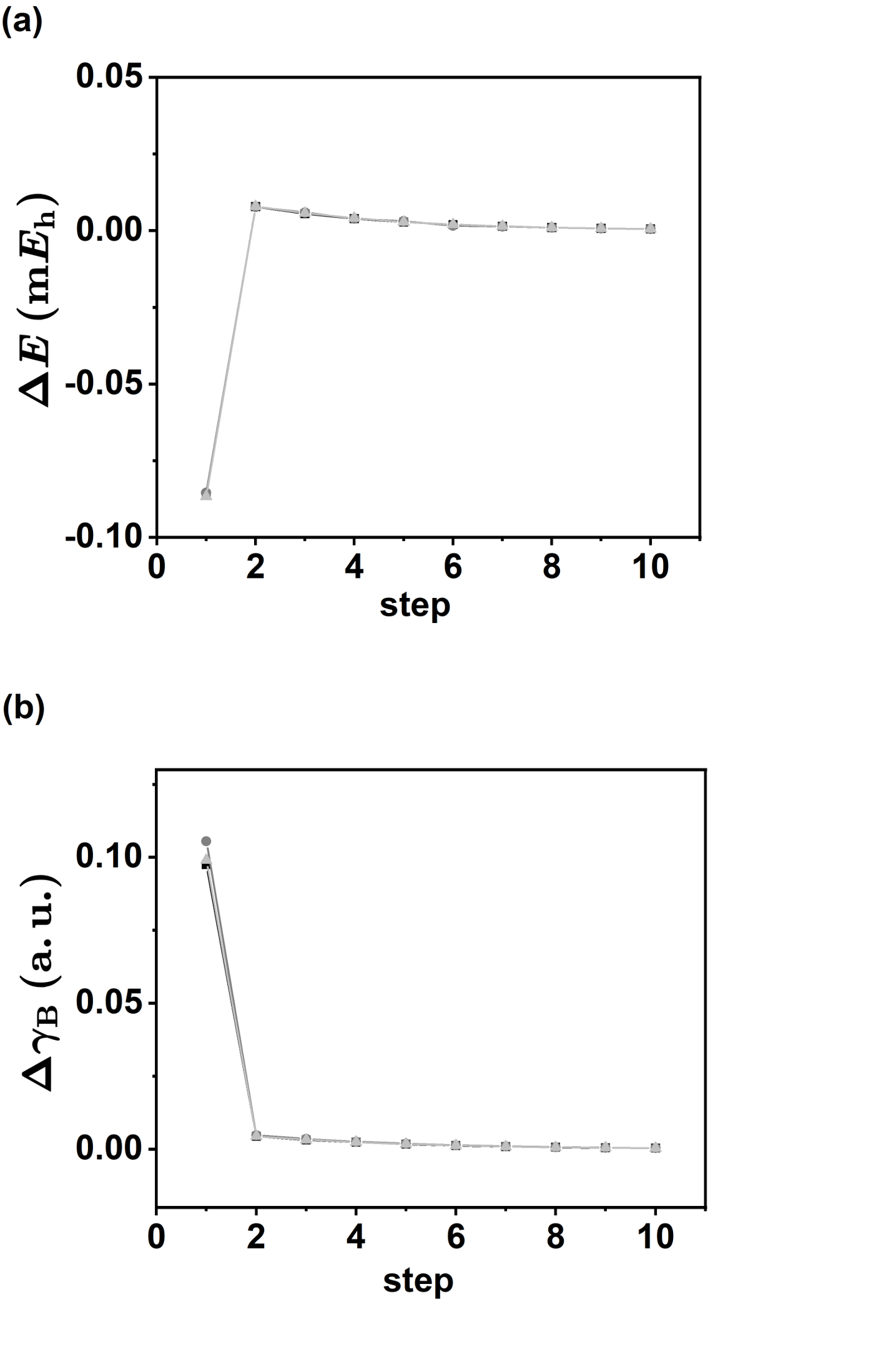}
    
\caption{Convergence of the iterative embedding procedure for an ethanol molecule with three molecular orbital localization schemes: (a) change in total energy at each macro-iteration step and (b) change in the environmental density. SPADE ($\blacksquare$), Boys ($\bullet$), and Pipek-Mezey ($\blacktriangle$) schemes yield nearly identical convergence profiles, demonstrating the robustness of the iterative procedure with respect to the choice of a localization scheme.}
\label{fig:ethanol_iteration_embedding}
\end{figure}

To further validate the robustness of the iterative procedure with respect to the choice of the MO localization scheme, we then employed an ethanol molecule, where subsystems A and B consisted of the hydroxyl and the ethyl groups, respectively. Again, subsystem A was treated at the VQE level while subsystem B was treated at the HF level. Figure \ref{fig:ethanol_iteration_embedding} displays the results with SPADE, Boys, and Pipek--Mezey localizations. Indeed, all three schemes yield virtually identical convergence profiles for both $\Delta E$ and $\Delta\gamma_\text{B}$, with only minor differences at the first iteration step. In addition, all three cases converged within $\sim$10 macro-iteration steps. This demonstrates that the iterative embedding procedure is robust with respect to the choice of the localization scheme and performs consistently across both non-covalently and covalently bonded systems.


    


\subsection{3.2 Improvement in Physical Description by Iterative Embedding}

While the results in the above establish the numerical robustness of the present scheme, the energy correction itself due to our iterative embedding remained rather small. 
This is because the two tested systems have limited amounts of static electron correlations, and thus the conventional embedding pictures before the macro-iterations were already good enough. 
To assess the benefit of adopting our scheme toward describing the physical aspects of any given system, let us now turn to a case where the electronic structure of the embedded subsystem undergoes a significant change by going from SCF to a higher level, as such a system may induce a stronger response in the environment density.

\begin{figure}[!ht]
    \centering
    \includegraphics[width=0.35\textwidth]{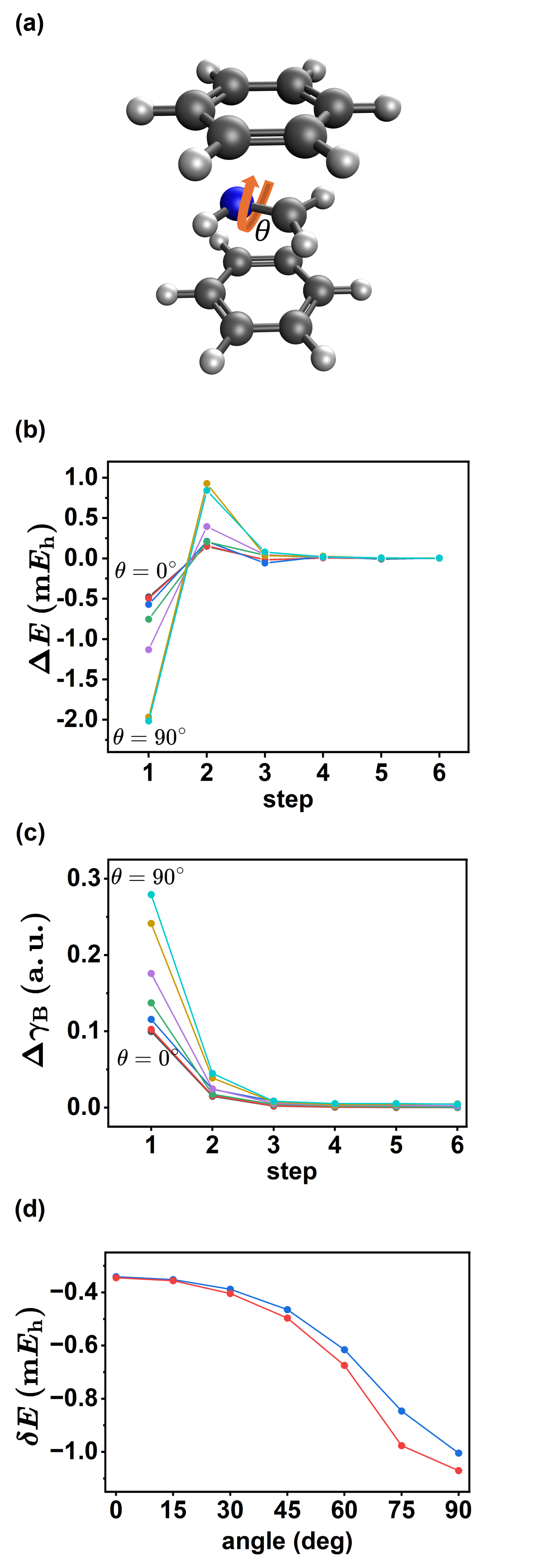}

\caption{(a) Molecular structure of the tested system consisting of a central methylenimine (CH$_2$NH) molecule sandwiched between two benzene rings. The twisted dihedral angle around C=N is shown with an arrow. (b) Convergence shown with the total energy change at each iteration step. (c) Convergence shown with the environmental density change at each iteration step. (d) Energy correction due to iterative embedding for VQE-in-HF (red) and CASCI-in-HF (blue) as a function of the dihedral angle.}
\label{fig:methylimine_iteration_embedding}
\end{figure}

To this end, we constructed a composite system with a methylenimine molecule (CH$_2$NH, subsystem A) sandwiched between two benzene rings (subsystem B), with the dihedral angle around the C=N bond sequentially twisted from 0$^\circ$ to 90$^\circ$ (Figure \ref{fig:methylimine_iteration_embedding}a), and treated it with the iterative VQE-in-HF scheme. This arrangement was specifically chosen based on the fact that the progressive disruption of the $\pi$-bond upon twisting will induce significant changes in the correlated electronic structure of the methylenimine subsystem, providing an ideal testbed for evaluating the response of the benzene environment. For MO localizations, the SPADE procedure was adopted. Considering the quantum resource constraints and computational costs associated with VQE, but at the same time to capture the essential electronic characteristics of the methylenimine moiety over the twisting, we employed a compact active space consisting of 3 occupied and 5 virtual orbitals. This led us to using 16 qubits for as many spin-orbitals, and the VQE simulations remained computationally feasible even with the iterative embedding burden.


Figure \ref{fig:methylimine_iteration_embedding}b presents the convergence of the stepwise energy difference between successive macro-iteration steps ($\Delta E = E_{i+1} - E_i$), evaluated across all rotation angles. Although the early iterations exhibit oscillatory behavior, reflecting the strong initial response of the benzene environment to the correlated electronic structure of methylenimine, the energies at all twisted geometries converged to their final answers within $\sim$5 macro-iteration steps. Figure \ref{fig:methylimine_iteration_embedding}c shows the corresponding convergence of the environmental density change, $\Delta\gamma_\text{B}$. The large values of $\Delta\gamma_\text{B}$ in the early stage also reflect the perturbation of the benzene environment induced by the updated correlated density of subsystem A, which in turn modifies the embedding potential and drives further changes in the embedded subsystem A. As the iterations proceeded, this feedback loop converged, and the environmental density reached steady values. The consistently converging behaviors across all dihedral angles with no geometry-dependent instability again demonstrate the numerical robustness of our iterative procedure.

Figure \ref{fig:methylimine_iteration_embedding}d shows the energy improvement $\delta E = E_\text{converged} - E_\text{initial}$, namely the difference between the iteratively converged VQE-in-HF energy ($E_\text{converged}$) and the initial VQE-in-HF energy ($E_\text{initial}$), as a function of the H$-$C=N$-$H dihedral angle. For comparison, we also present the results obtained by employing the complete active space configuration interaction (CASCI) method for subsystem A, namely the CASCI-in-HF results. At all angles, $\delta E$ is negative, indicating that the iterative procedure consistently lowers the total energy relative to the original frozen embedding, and the magnitude of $\delta E$ grows monotonically with the dihedral angle, reaching approximately $-$1.0 $\mathrm{m}E_\mathrm{h}$ at 90$^\circ$. This trend reflects the increasing importance of the mutual polarization between the twisted methylenimine and the benzene enenvironment as the C=N $\pi$-bond is progressively disrupted. The close agreements between VQE-in-HF and CASCI-in-HF across all angles demonstrate that VQE reliably reproduces the CAS reference within the same active space, validating that the improvement from the iterative embedding indeed has a physical origin.

\begin{figure}[!ht]
    \centering
    \begin{subfigure}{0.45\textwidth}
        \centering
        \includegraphics[width=\textwidth]{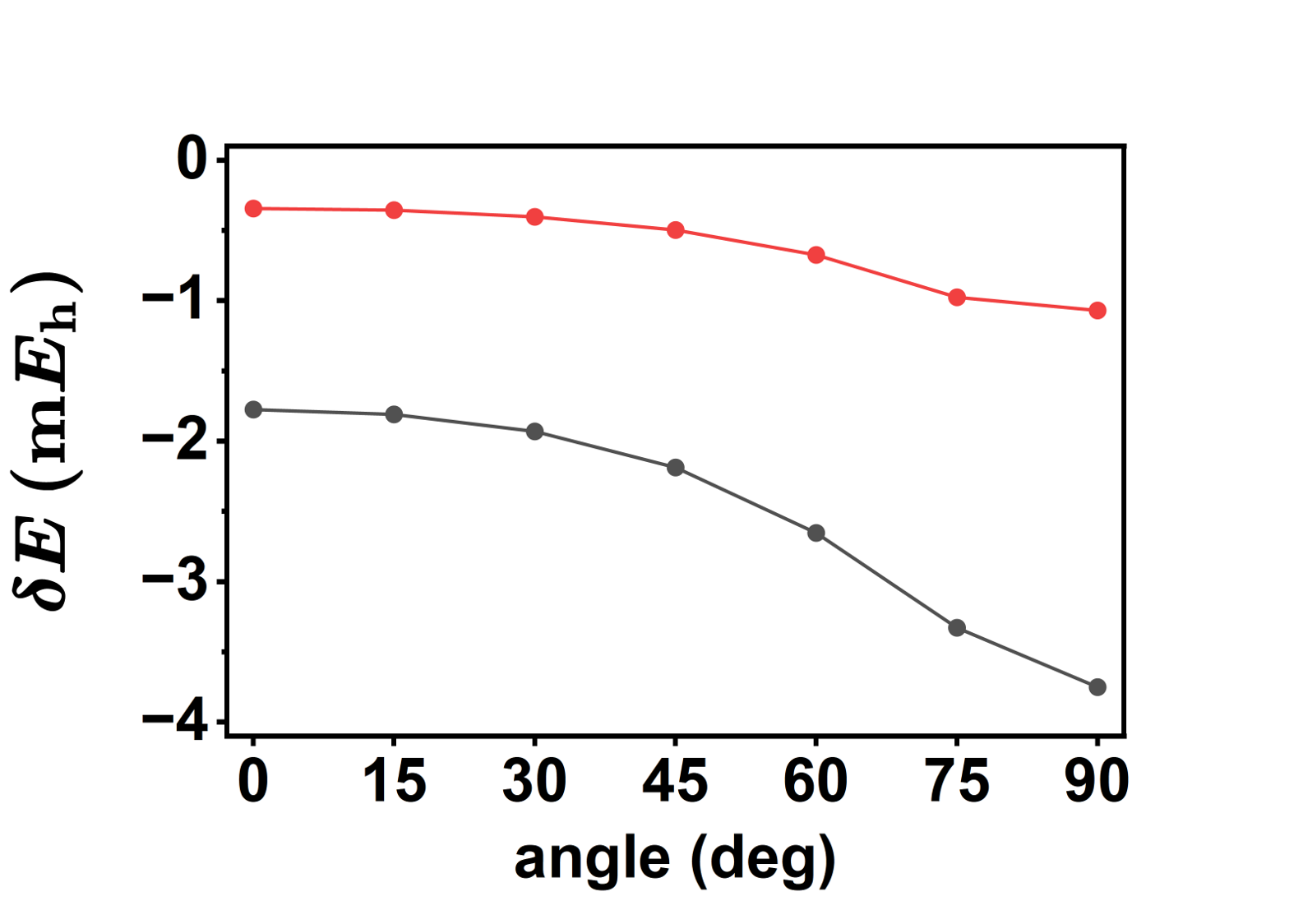}
    \end{subfigure}\\[-0.5em] 

\caption{Comparison of non-iterative to iterative energy correction with VQE-in-HF (red) versus CASCI to CASSCF energy correction (black) with CH$_2$NH at different H$-$C=N$-$H dihedral angles.}
\label{fig:methylimine_iteration_embedding2}
\end{figure}

As a next step, we considered it prudent to determine how much correlation energy we were recovering from our iterative scheme. For this purpose, we inspected the energy lowering gained by employing CASSCF in the place of whole-system CASCI and compared it against the energy lowering from our iterative scheme.
Of course, care must be taken in directly comparing our results with CASSCF ones, as our scheme does not employ the same orbital relaxation mechanism as in CASSCF. For example, UCCSD-based VQE cannot mix predefined active and inactive orbitals, while CASSCF can. Nevertheless, the comparison can still be meaningful for assessing the upper bound of energy lowering that can be attained by the orbital relaxation. For the
comparison, orbital optimization was carried out from the CASCI-in-HF result by allowing exchanges of active orbitals and unoccupied orbitals, spanning over both subsystems A and B. Doubly occupied core orbitals were kept frozen to avoid potential irregularity associated with our relatively small active space definition, which was dictated by the resource limitation in the VQE simulations. Figure \ref{fig:methylimine_iteration_embedding2} compares the energy lowering obtained from our iterative embedding procedure with that from switching from CASCI to CASSCF. Across the entire range of dihedral angles, the correction gained from the iterative embedding amounts to approximately 20--30\% of the CASCI-to-CASSCF energy lowering. Furthermore, both quantities become increasingly negative as the dihedral angle increases, indicating that the importance of relaxation grows as the C=N $\pi$-bond is progressively disrupted. These results suggest that the proposed self-consistent embedding procedure captures a significant fraction of the correlation-driven relaxation effects associated with the development of multireference character.

\subsection{3.3 Fictitious Energy by the Projection Penalty}

\begin{figure}[ht]
\begin{subfigure}{0.60\textwidth}
\centering
\includegraphics[width=\textwidth]{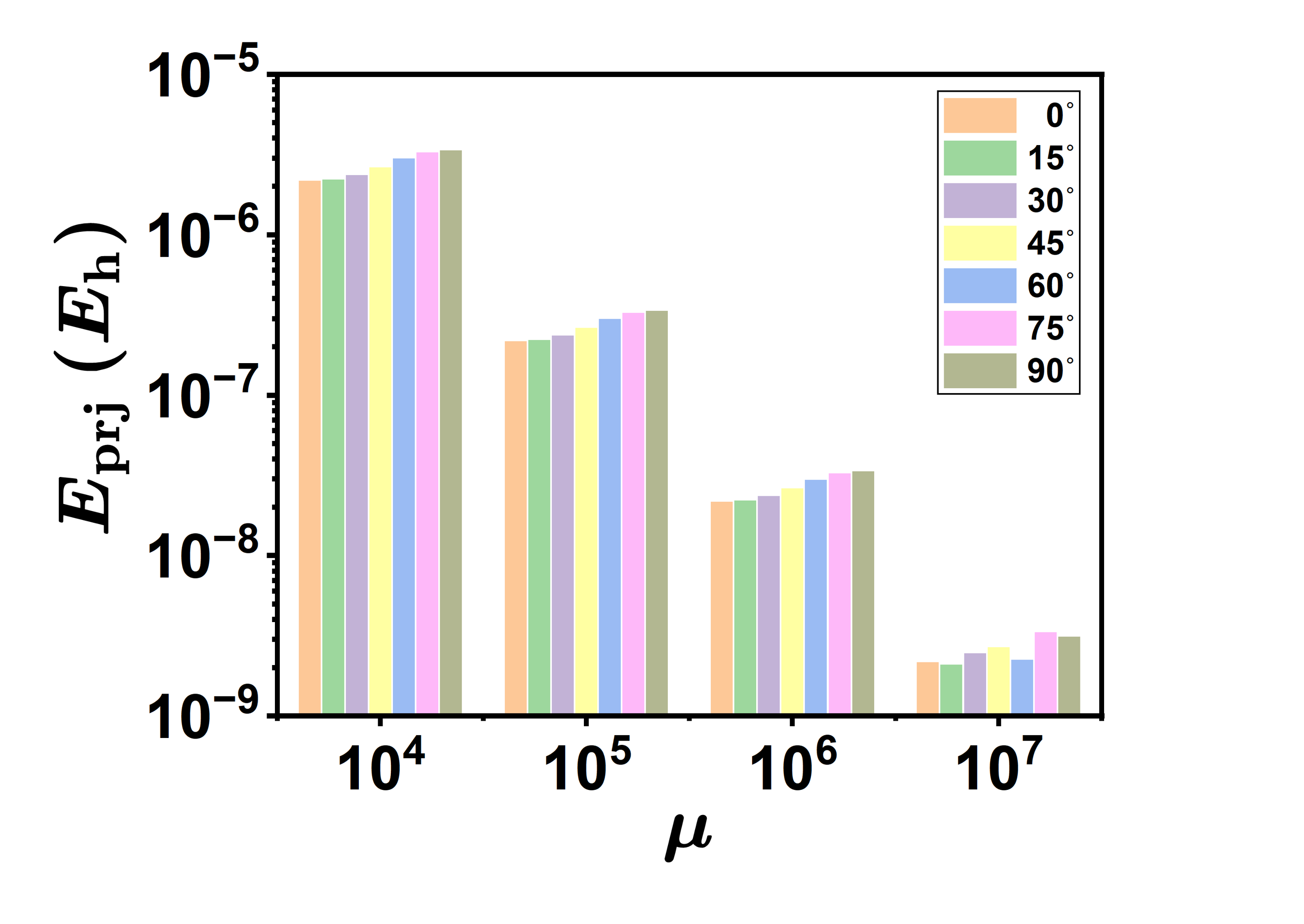}
\end{subfigure}
\caption{Fictitious energy from the projection penalty
as a function of $\mu$ (horizontal axis) and dihedral angle (color-coded).}
\label{fig:level_shift_contribution}
\end{figure}

The projection operator employed in the projection-based embedding is formally defined using the reference SCF density, which corresponds to a single Slater determinant. Therefore, it defines a unique orbital space formed by occupied orbitals of the environment. In the present iterative scheme, however, the subsystem density used during the B-in-A update (Step 2) is obtained from VQE and therefore represents a non-single determinant density matrix. This density is related to a multi-configurational wavefunction and does not, in general, define a unique occupied orbital space. In fact, the projection operator based on this VQE density performs weighted projection\cite{mayer2014effective} on the natural orbitals that correspond to the VQE density, with the weight of each orbital reflecting its extent of occupation. The mutual orthogonality designed in the original embedding scheme will still be enforced by the subsequent A-in-B calculation during the next macro-iteration. Even still, the penalty energy term $E_\text{prj}=\mu \, \mathrm{Tr} \gamma^\text{B} \mathbf{S} \gamma^\text{A}_\text{VQE} \mathbf{S}$ included eventually in Eq.~(\ref{eq:E_WFT-in-SCF}) is no longer guaranteed to vanish exactly, and it will be a discreet test to check whether this aspect would cause any trouble. Figure \ref{fig:level_shift_contribution} shows this penalty energy contribution as a function of the level-shift parameter $\mu$ for the case of the CH$_2$NH--(benzene)$_2$ complex. The contribution decreases systematically with increasing $\mu$, reaching approximately $10^{-6}$ a.u. for $\mu=10^{4}$ and $10^{-9}$ a.u. for $\mu=10^{7}$. As all molecular calculations in this work have employed $\mu=10^{7}$, the penalty contribution will indeed be negligible compared to the total electronic energy. Indeed, the results indicate that the use of the VQE density in the projection operator for the iteration will not introduce any practically significant error.

\section{4. Conclusion}

In this work, we have developed an iterative projection-based embedding framework that combines VQE with a self-consistent environment relaxation scheme, going beyond the conventional one-shot embedding approximation. Rather than keeping the environment density frozen at the initial SCF solution, the proposed method allows the environment to respond iteratively to the updated electronic structure of the embedded subsystem through a macro-iteration cycle consisting of three sequential steps: a conventional A-in-B calculation, a reversed B-in-A SCF refinement with damped A-density, and a reconstruction of the embedded Hamiltonian.

The numerical robustness of the iterative procedure was validated on two systems with distinct interaction characters: a water dimer and an ethanol molecule. For the water dimer, convergence to a unique total energy was achieved within $\sim$10 macro-iteration steps across all tested damping factors, demonstrating consistency regardless of the strength of damping. For ethanol, for which the two subsystems were divided across a covalent bond, the procedure again converged in a practically identical pattern with three different localization schemes (SPADE, Boys, and Pipek--Mezey), confirming that the method is robust with respect to the choice of a molecular orbital partitioning scheme.

The physical impact of the iterative environment relaxation was assessed on a methylenimine molecule sandwiched between two benzene rings, with the dihedral angle around C$=$N twisted from 0$^\circ$ to 90$^\circ$. Our iterative embedding consistently converged within 5--6 macro-iteration steps at all twisted geometries. The converged energies were consistently lower than those from the initial frozen embedding, with the energy correction growing monotonically with the dihedral angle. This reflects the increasing response of the benzene environment as the $\pi$-bond was progressively disrupted. In addition, it shows that our iterative embedding scheme can indeed be useful when characterizing a system with a strongly-correlated subsystem whose electronic structure deviate much from its mean-field level description. 

Looking forward, extending projection-based embedding to correlated wavefunctions remains an important challenge. The present work shows that the correlated density introduces only negligible violation of the orthogonality condition for the molecular systems considered here. We hope that our formulation of a projection scheme will be rigorously applicable to obtaining correlated densities even in larger systems in general, to provide a stronger and broader applicability of the projection-based embedding with quantum hardware in future studies.

\begin{acknowledgement}
This work was supported by the National Research Foundation of Korea (NRF) grants funded by the Korea government (MSIT), with Nos. RS-2020-NR049542, RS-2023-NR119931, and RS-2024-00432113. Fruitful discussions with Young Kyun Ahn are also gratefully acknowledged. 
\end{acknowledgement}


\bibliography{achemso-demo}

\end{document}